\documentclass[12pt]{iopart}
\usepackage{iopams}  
\usepackage{epsfig}
\usepackage{graphicx}

\begin{document}

\title[$\langle m_T \rangle$ excitation function]{$\langle m_T \rangle$ excitation function: Freeze-out and equation of state dependence}

\author{Hannah Petersen${}^{1,2}$, Jan Steinheimer${}^2$, Marcus Bleicher${}^2$ and Horst St\"ocker${}^{1,2,3}$\\[.4cm]}

\address{${}^1$~Frankfurt Institute for Advanced Studies (FIAS), Ruth-Moufang-Str.~1, D-60438 Frankfurt am Main,
Germany\\
${}^2$~Institut f\"ur Theoretische Physik, Johann Wolfgang Goethe-Universit\"at, Max-von-Laue-Str.~1, 60438 Frankfurt am Main, Germany \\
${}^3$~GSI Helmholtzzentrum f\"ur Schwerionenforschung (GSI), Planckstr.~1, D-64291 Darmstadt, Germany}

\ead{petersen@th.physik.uni-frankfurt.de}

\begin{abstract}
An integrated Boltzmann+hydrodynamics transport approach is applied to investigate the dependence of the mean transverse mass on the freeze-out and the equation of state over the energy range from $E_{\rm lab}=2-160A~$GeV. This transport approach based on the Ultra-relativistic Quantum Molecular Dynamics (UrQMD) with an intermediate hydrodynamic stage allows for a systematic comparison without adjusting parameters. We find that the multiplicities of pions and protons are rather insensitive to different freeze-out prescriptions and changes in the equation of state, but the yields are slightly reduced in the hybrid model calculation compared to a pure transport calculation while the (anti)kaon multiplicities are increased. The mean transverse mass excitation functions of all three particle species are found to be sensitive to the different freeze-out treatments as well as to the equation of state. We find that the bag model equation of state with a strong first order phase transition is in qualitative agreement with the experimentally observed step-like behaviour in the $\langle m_T \rangle$ excitation function. The hybrid model with a hadron gas equation of state leads to a overestimation of the $\langle m_T \rangle$, especially at higher energies. However, non-equilibrium effects seem also to be substantial as is suggested by the comparison with standard UrQMD results.   
\end{abstract}


\maketitle

\section[Introduction]{Introduction}
\label{intro}
To explore the deconfinement phase transition from a hadron gas to the so called Quark Gluon Plasma (QGP) is one of the main motivations to study heavy ion collisions at high energies \cite{Stoecker:1986ci}. The mean transverse momentum excitation function has been proposed as a signal for the observation of this phase transition many years ago \cite{VanHove:1982vk} and has been further explored in the following years \cite{Scherer:1999qq,Xu:2001zj,Gorenstein:2003cu,Gazdzicki:2003dx,Cleymans:2007uk}. Experimentally, the energy dependences of various observables (including the mean transverse mass $\langle m_T \rangle$) show anomalies at low SPS energies which might be related to the onset of deconfinement \cite{Gazdzicki:1998vd,Gazdzicki:2004ef,:2007fe}.

The mean transverse mass of the particles is expected to be proportional to the temperature at kinetic freeze-out of the particles. If there is a phase transition the mean transverse mass excitation function should show a flattening because of the mixed phase region where the temperature of the system stays constant. However, recently, it has been suggested in a 3-fluid model with a hadron gas equation of state that the characteristic steplike-behaviour of the mean transverse mass in dependence of the incident beam energy might be a relic of the freeze-out treatment in a hydrodynamic calculation and may not be a unique sign of the transition to the QGP \cite{Ivanov:2006cz}.  

To overcome this ambiguity we suggest here to employ an integrated Boltzmann+ hydrodynamics approach with an explicit microscopic treatment of the freeze-out procedure and different equations of state. These microscopic+macroscopic approaches have been very successful in describing many features of heavy ion collisions. The explicit inclusion of a hydrodynamic part has the great advantage that the equation of state is a well-defined input to the calculation and that the phase transition can be easily treated. In addition, using a transport approach for the initial conditions and the hadronic interactions after the freeze-out from the hydrodynamic stage has already been shown to be a promising idea and is currently considered as state-of-the-art for the exploration of hot and dense QCD matter \cite{Nonaka:2006yn,Dumitru:1999sf,Bass:2000ib,Teaney:2001av,Grassi:2005pm,Andrade:2005tx,Hirano:2005xf,Hirano:2007ei,Petersen:2008dd,Li:2008qm}. The main advantage of these kinds of calculations is that the initial state and the freeze-out conditions can be fixed by only two parameters for the whole energy range. Then, the only remaining input is the equation of state (EoS). Therefore, this approach provides a particularly well suited basis to extract information about effects of different EoS on observables. As a baseline check, purely hadronic transport and hydrodynamic calculations can be compared in the same framework to explore viscosity effects and changes in the dynamics without altering the degrees of freedom.  

Here, we investigate the effects of different freeze-out implementations and changes in the equation of state (EoS) in an integrated (3+1)dimensional Boltzmann+hydrodynamics approach. This paper is structured in the following way. In the Section \ref{model} a short description of the applied hybrid approach is given. Sections \ref{freezeout} and \ref{eos} contain the results for multiplicities and the mean transverse mass of pions, kaons and protons from AGS to SPS energies where we first concentrate on the dependence on the freeze-out prescription and then explore three different equations of state. Section \ref{sum} summarizes the paper.  

\section[Model Description]{Model Description}
\label{model}

The present hybrid model used to simulate the dynamics of heavy ion collisions from $E_{\rm lab}=2-160A~$GeV is based on the UrQMD approach \cite{Bass:1998ca,Bleicher:1999xi} with an intermediate ideal hydrodynamic evolution for the hot and dense stage of the reaction. The advantages of both approaches are combined in such a way that no additional adjustments of parameters is necessary for different energies or centralities. To mimic experimental conditions as realistically as possible the non-equilibrium dynamics in the initial and the final state are taken into account on an event-by-event basis (see e.g. discussions in \cite{Bleicher:1998wd,Grassi:2005pm,Andrade:2005tx,Andrade:2006yh,Tavares:2007mu,Andrade:2008xh}). 

UrQMD is a string/hadronic transport approach which simulates multiple interactions of ingoing and newly produced particles, the excitation and fragmentation of color strings \cite{NilssonAlmqvist:1986rx,Sjostrand:1993yb} and the formation and decay of hadronic resonances. In equilibrium, UrQMD exhibits a hadron gas equation of state \cite{Bravina:1998it,Bravina:1998pi,Belkacem:1998gy}. The coupling between the UrQMD initial state and the hydrodynamical evolution proceeds when the two Lorentz-contracted nuclei have passed through each other, at $t_{\rm start} = {2R}/{\sqrt{\gamma^2-1}}$ \cite{Steinheimer:2007iy}. This leads to non-trivial velocity and energy density distributions for the hydrodynamical initial conditions \cite{Petersen:2009vx}. 

Starting from these (single event) initial conditions a full (3+1) dimensional ideal hydrodynamic evolution is performed using the SHASTA algorithm \cite{Rischke:1995ir,Rischke:1995mt}. The hydrodynamic evolution is stopped if a certain transition criterion is fulfilled. Here, we investigate two different commonly employed conditions: 

\begin{itemize}
\item Isochronous freeze-out (IF): 

The hydrodynamic evolution is stopped, if the energy density $\varepsilon$ of all cells drops below five times the ground state energy density $\varepsilon_0$ (i.e. $\sim 730 {\rm MeV/fm}^3$). This criterion is fixed for all beam energies and equations of state. In Section \ref{freezeout} we show the effects of a change of the $\epsilon_0$ parameter on the results. The hydrodynamic fields are then mapped to particle degrees of freedom via the Cooper-Frye equation on an isochronous (at the computational frame time $t_{\rm CM}$) hypersurface. The particle vector information is then transferred back to the UrQMD model, where rescatterings and final decays are calculated using the hadronic cascade. We will further refer to this kind of freeze-out procedure as the ischronous freeze-out (IF).

\item Iso-eigentime freeze-out (GF): 

In \cite{Li:2008qm} we have introduced another freeze-out procedure to account for the large time dilatation that occurs for fluid elements at large rapidities. To mimic an iso-$\tau$ hypersurface we freeze out full transverse slices, of thickness $\Delta z = 0.2 $fm, whenever all cells of that slice fulfill the freeze-out criterion. For each individual slice the isochronous (in $t$) Cooper-Frye procedure as described above is applied. Doing this one obtains a rapidity independent freeze-out temperature even for the highest beam energies. In the following we will refer to this procedure as ``gradual freeze-out''(GF). 

\end{itemize}

For a detailed description of the hybrid model including parameter tests and results for multiplicities and spectra the reader is referred to \cite{Petersen:2008dd}.

Serving as an input for the hydrodynamical calculation the EoS strongly influences the dynamics of an expanding system. In this work we use the following three different equations of state that are introduced in \cite{Li:2008qm} in detail:

\begin{itemize}
\item a hadron gas EoS describing a non-interacting gas of free hadrons with masses up to 2 GeV (HG)\cite{Zschiesche:2002zr},
\item a bag model EoS that exhibits a strong first order phase transition between a Walecka type hadron gas and massless quarks and gluons (BM) \cite{Rischke:1995mt},
\item a chiral EoS that follows from a chiral hadronic SU(3) Lagrangian that is able to reproduce a phase diagram structure including a moderate first order phase transition at high-$\mu_B$ values and a critical endpoint \cite{Papazoglou:1998vr,Zschiesche:2006rf,Steinheimer:2007iy} (CH).
\end{itemize}

With these ingredients we employ the present implementation of the model to disentangle different effects like the change of the dynamics between hydrodynamics and transport theory and freeze-out prescriptions in a consistent manner, thus providing the opportunity to explore different equations of state within the same framework. 

\section[Freeze-out Dynamics]{Freeze-out Dynamics}
\label{freezeout}

\begin{figure}[t]
\includegraphics[width=0.5\textwidth]{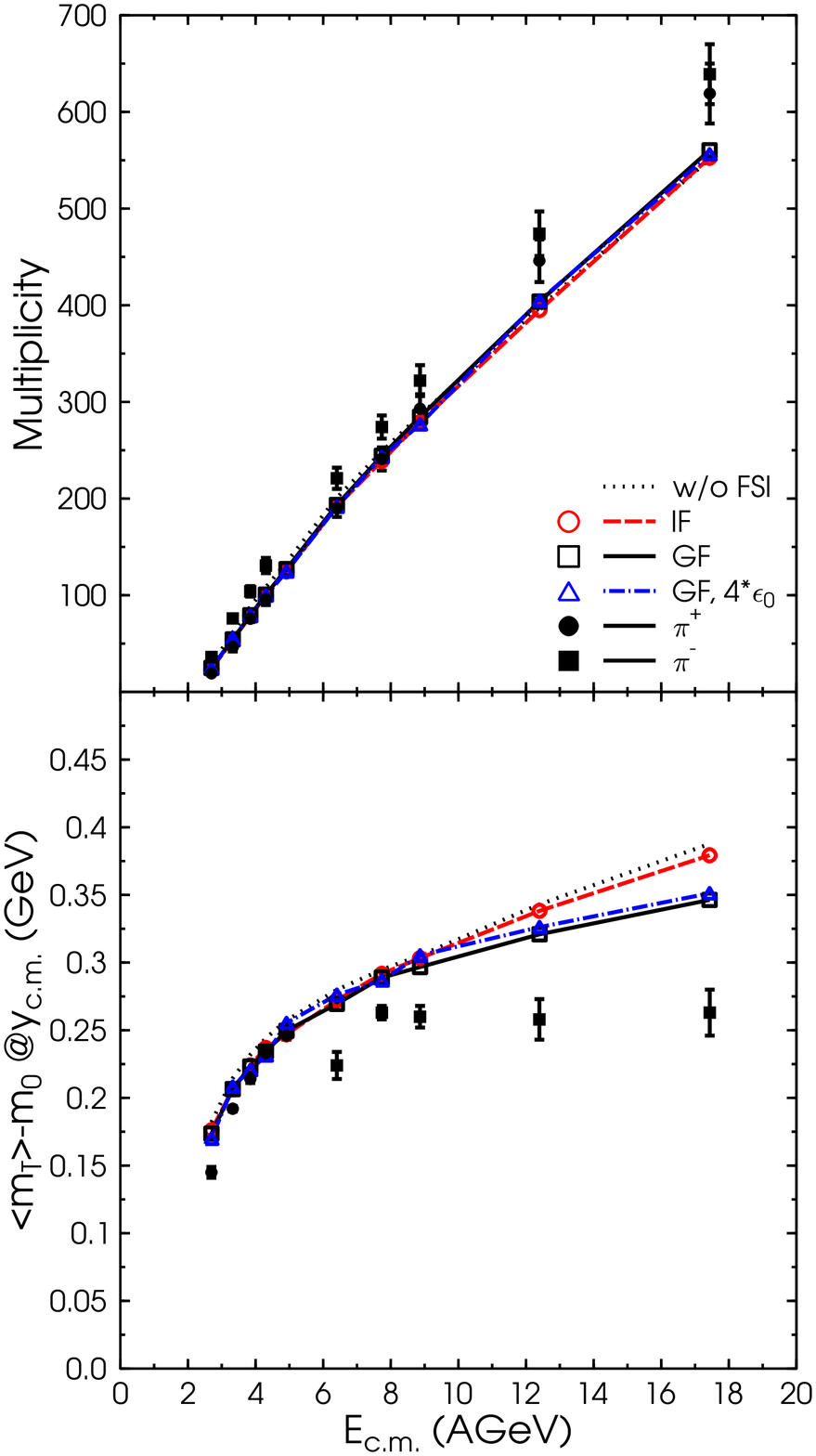}
\includegraphics[width=0.5\textwidth]{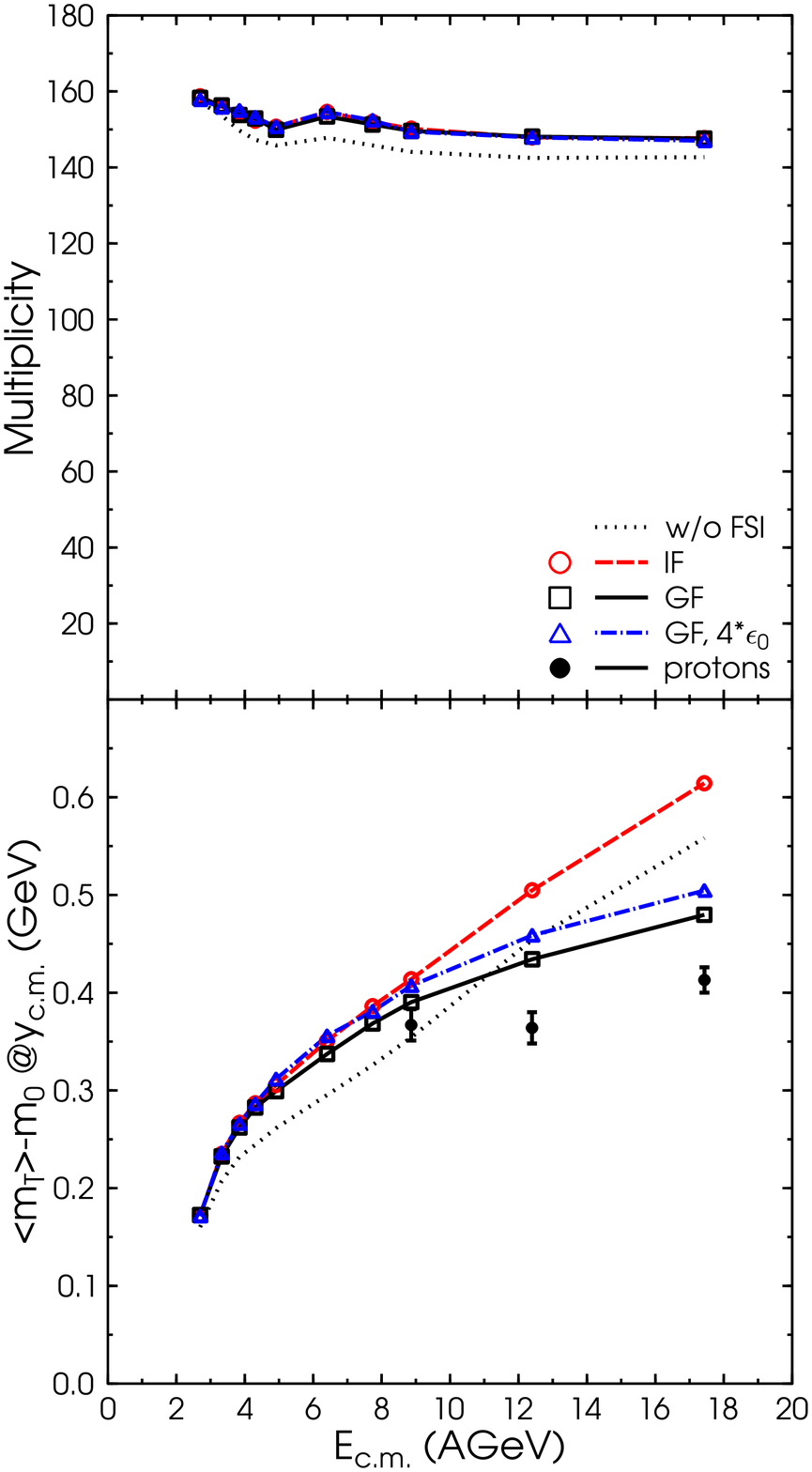}
\caption[Freeze-out dependence of the $\langle m_T \rangle$ excitation function of $\pi$'s and protons]{\label{fig_fopi} The multiplicity (4$\pi$, top) and the $\langle m_T \rangle$ (bottom) excitation function for pions (left) and protons (right) in central ($b<3.4$ fm) Au+Au/Pb+Pb collisions at $E_{\rm lab}=2-160 ~A$GeV is shown. The lines depict different freeze-out prescriptions for the hybrid model calculation with the hadron gas equation of state, while the symbols depict experimental data \cite{:2007fe,Afanasiev:2002mx,Ahle:1999uy,Anticic:2004yj}.}
\end{figure}

\begin{figure}[t]
\includegraphics[width=0.5\textwidth]{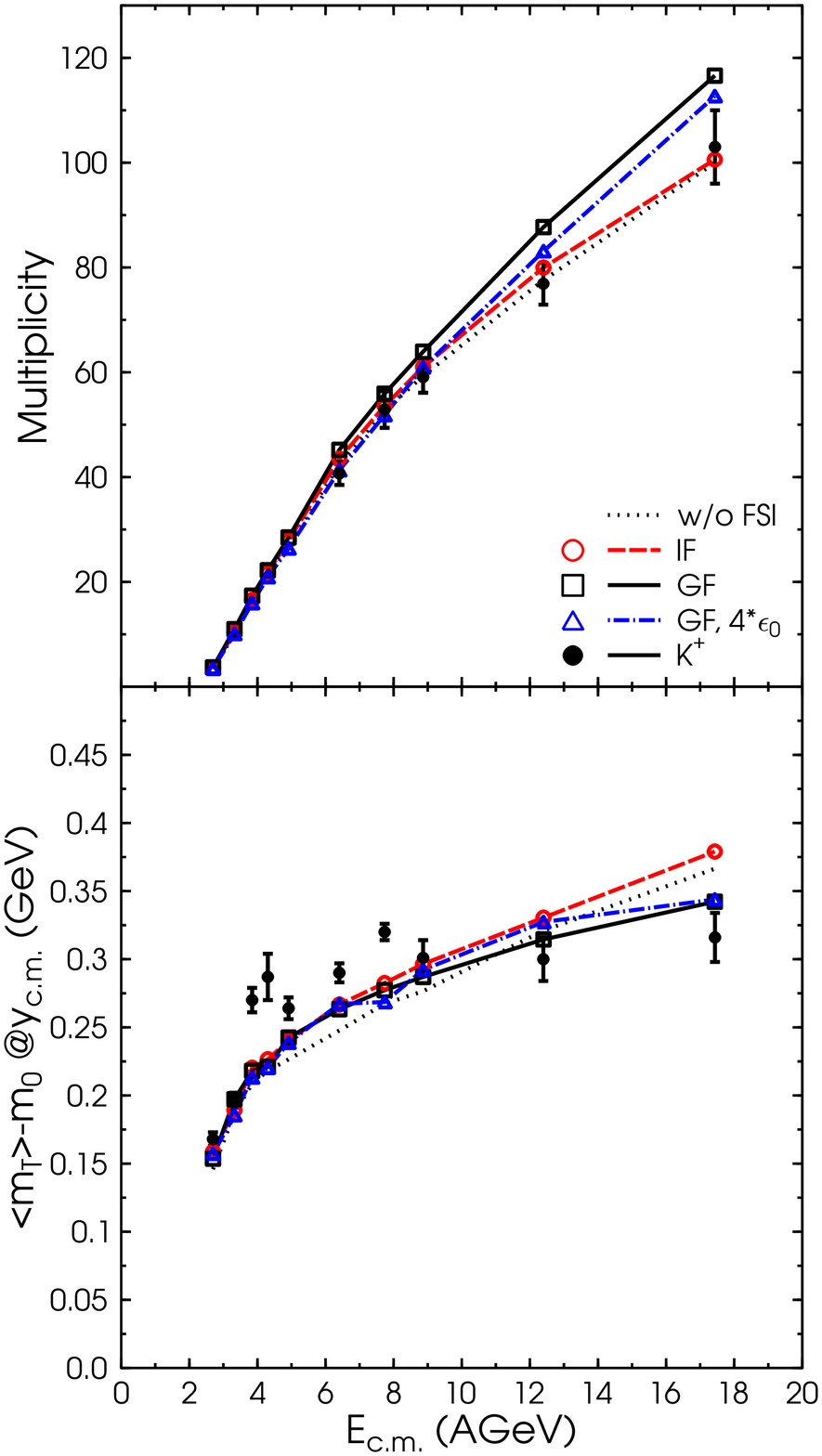}
\includegraphics[width=0.5\textwidth]{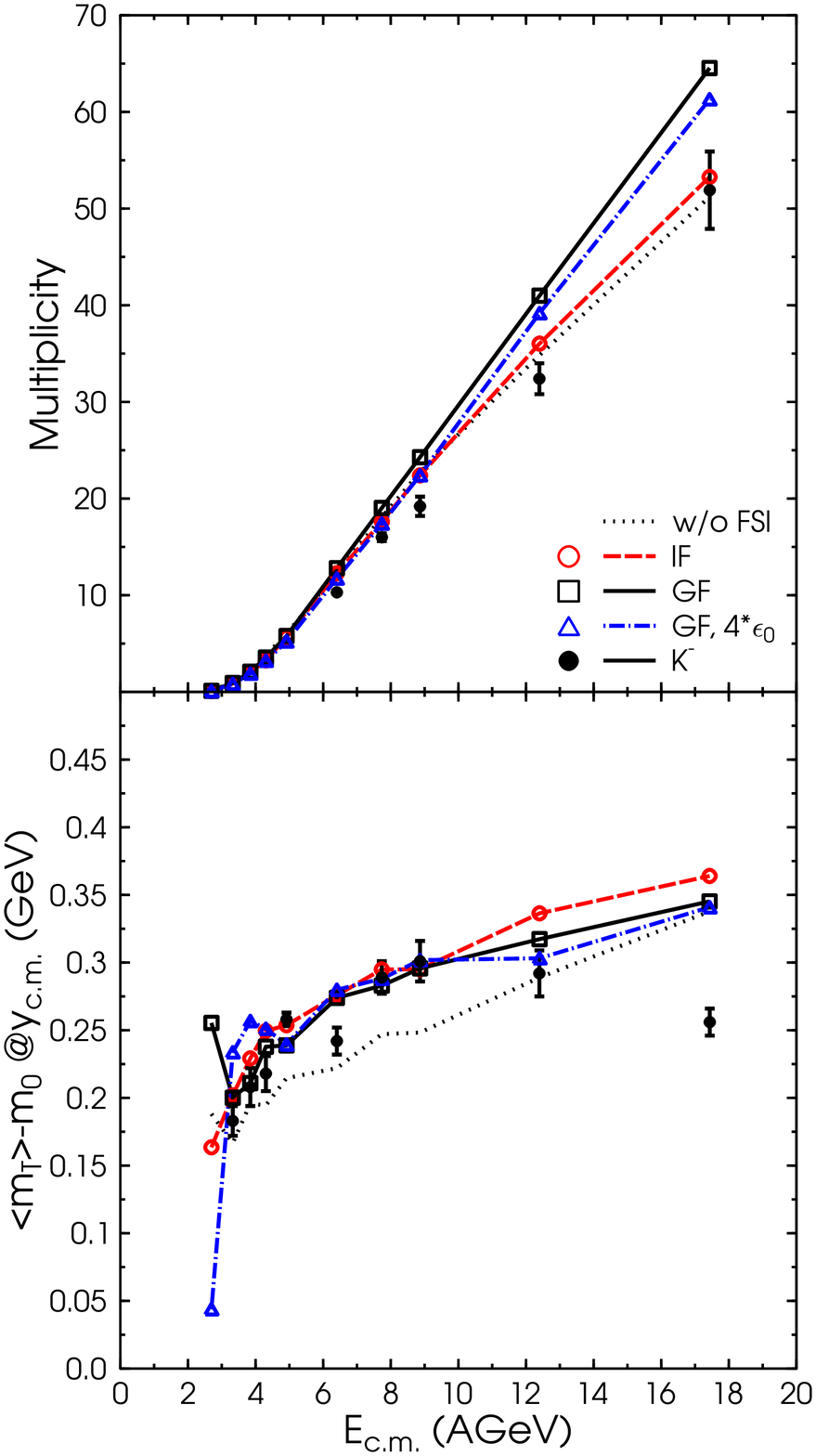}
\caption[Freeze-out dependence of the $\langle m_T \rangle$ excitation function of kaons]{\label{fig_fokp} The multiplicity (4$\pi$, top) and the $\langle m_T \rangle$ (bottom) excitation function for positively/negatively (left/right) charged kaons in central ($b<3.4$ fm) Au+Au/Pb+Pb collisions at $E_{\rm lab}=2-160 ~A$GeV is shown. The lines depict different freeze-out prescriptions for the hybrid model calculation with the hadron gas equation of state, while the symbols depict experimental data \cite{:2007fe,Afanasiev:2002mx,Ahle:1999uy,Ahle:2000wq}. }
\end{figure}

In this Section we address the $4\pi$ multiplicity (Figs. \ref{fig_fopi}-\ref{fig_fokp}, upper plots) and the mean transverse mass (Figs. \ref{fig_fopi}-\ref{fig_fokp}, lower plots) excitation functions for pions, protons and kaons calculated in the hybrid approach with a hadronic EoS to compare different freeze-out prescriptions. The dotted line corresponds to the results directly after the isochronous hydrodynamic freeze-out without final state interactions (FSI). Immediate decay of the resonances is taken into account to provide comparable multiplicity results. All other calculations include the hadronic afterburner, however with different transition prescriptions applied: the isochronous freeze-out (dashed line with circles), the gradual freeze-out (full line with squares) and the gradual freeze-out with varied freeze-out energy density criterion $4 \epsilon_{0}$ (dashed-dotted line with triangles). 

In the following figures the beam energy dependence of the multiplicities (top) and the mean transverse mass (bottom) is shown. The results are calculated for central Au+Au/Pb+Pb collisions at midrapidity ($|y|<0.5$) from $E_{\rm lab}=2-160A~$GeV. In general, one observes that the mean transverse mass increases as a function of energy, because more energy becomes available that can be distributed in the transverse plane and the expansion phase lasts longer. 

Let us start the detailed discussion with the pions as being the most abundant hadrons. The pion multiplicity (Fig. \ref{fig_fopi}, left) is completely insensitive to the freeze-out procedure while the mean transverse mass at higher energies is decreased if the gradual freeze-out procedure is applied. The final state interactions and the variation of the freeze-out criterion do  only weakly alter the results for pions. At AGS energies, the calculations are well in line with the data while at SPS energies where the hydrodynamic stage is a major part of the evolution the pion multiplicity stays below the data and the mean transverse mass exceeds the experimental data. We attribute these observations to the entropy conservation in the hydrodynamic evolution and the violent transverse expansion because of high pressure gradients.   

Next, we explore the production and expansion of the baryon charge. In contrast to the previous case, the proton multiplicity (Fig. \ref{fig_fopi}, right) is almost constant over the whole energy range. The final rescatterings lead in this case to a slightly higher multiplicity and an increased mean transverse mass. The protons acquire more transverse flow during the hadronic stage after the hydrodynamic evolution. As already observed for the pions the gradual freeze-out leads to a flattening of the transverse mass excitation function. Varying the energy density criterion (squares vs triangles) indicates only a weak dependence on this parameter.

Finally, we address strange particles. Fig. \ref{fig_fokp} shows the results for positively and negatively charged kaons respectively. The kaon multiplicities are nicely reproduced, if the isochronous freeze-out procedure is applied with or without final state interactions. The mean transverse mass in the same calculation is too high at top SPS energies due to the violent transverse expansion because of the comparably stiff hadronic EoS. For $K^-$ the final rescatterings lead to an increase in the mean transverse mass because the low $p_T$ $K^-$ are easily absorbed on the surrounding baryons. In the gradual freeze-out scenario the kaon production is enhanced by roughly 10\% due to the higher average temperatures at the transition point from hydrodynamics to the transport description at higher energies. For the kaon multiplicity the variation of the freeze-out criterion leads to a slight decrease of the yields when going from the gradual to the isochronous scenario. This reflects the sensitivity of the kaon yield to the temperature at the transition point between hydrodynamics and the final state hadron cascade. The mean transverse mass excitation functions are flatter with the gradual freeze-out scenario which leads to a better description of the experimental data.

Overall, the hybrid calculation with the gradual freeze-out procedure reproduces the multiplicities and the shape of the mean transverse mass excitation function best, in many cases there is even good quantitative agreement with the experimental data. 

\section[EoS Dependence]{Equation of State Dependence}
\label{eos}

After these rather technical studies, we turn now to the investigation of different EoS. To exemplify the effects of the different underlying dynamics we contrast the hybrid model calculations with the pure hadronic transport calculation (UrQMD-2.3, dotted line). By comparing this calculation with the hybrid calculation (employing the HG EoS, full line with squares) one can estimate the effect of viscosities and the non-equilibrium dynamics. For the hybrid model calculations the gradual freeze-out with the criterion of $5\epsilon_{0}$ is applied as it provides the best fit to the data as shown before. The dashed-dotted line with triangles corresponds to the calculation with the chiral equation of state (CH) while the dashed line with circles depicts the bag model EoS with a strong first order phase transition (BM). 

\begin{figure}[t]
\includegraphics[width=0.5\textwidth]{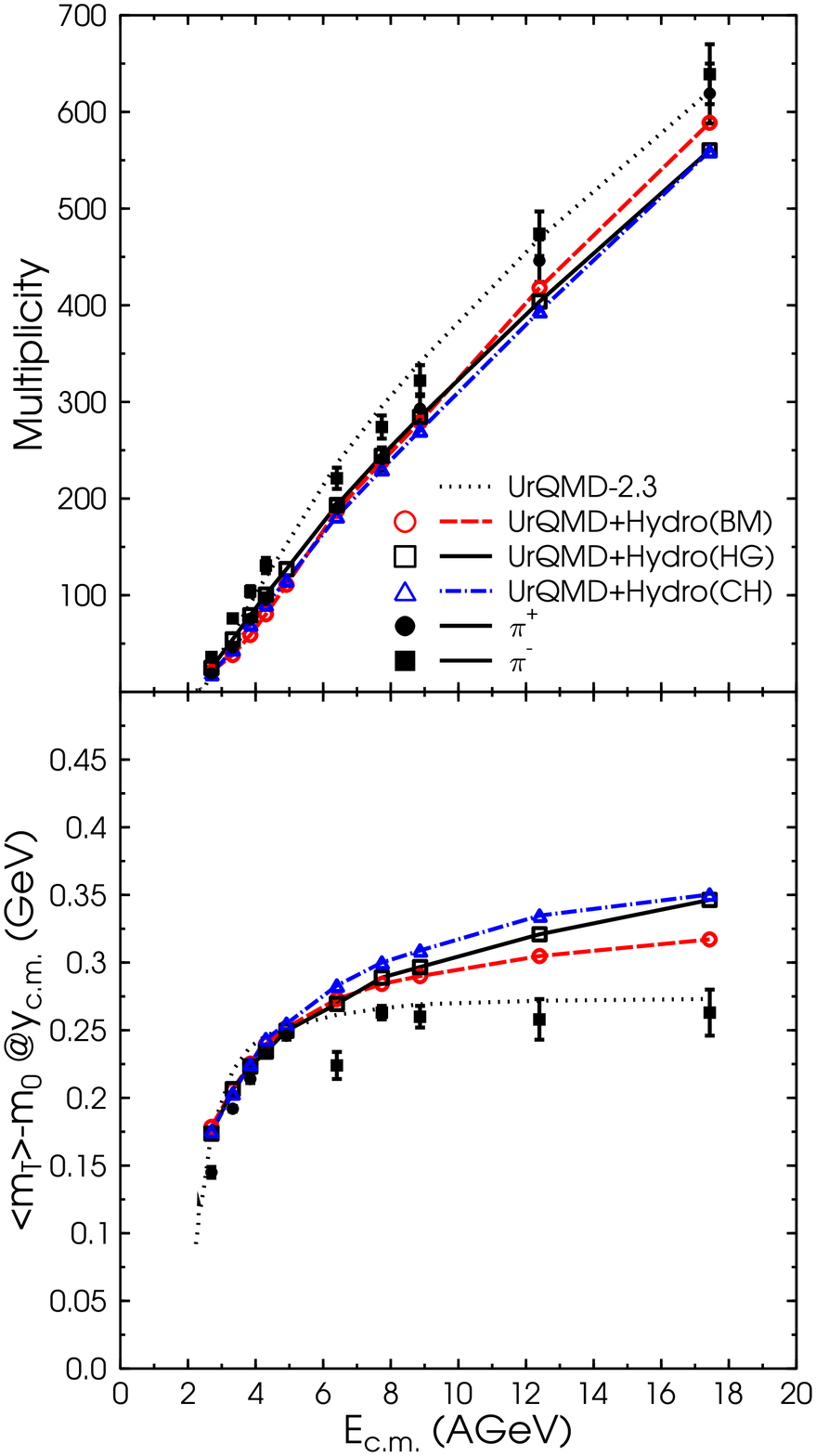}
\includegraphics[width=0.5\textwidth]{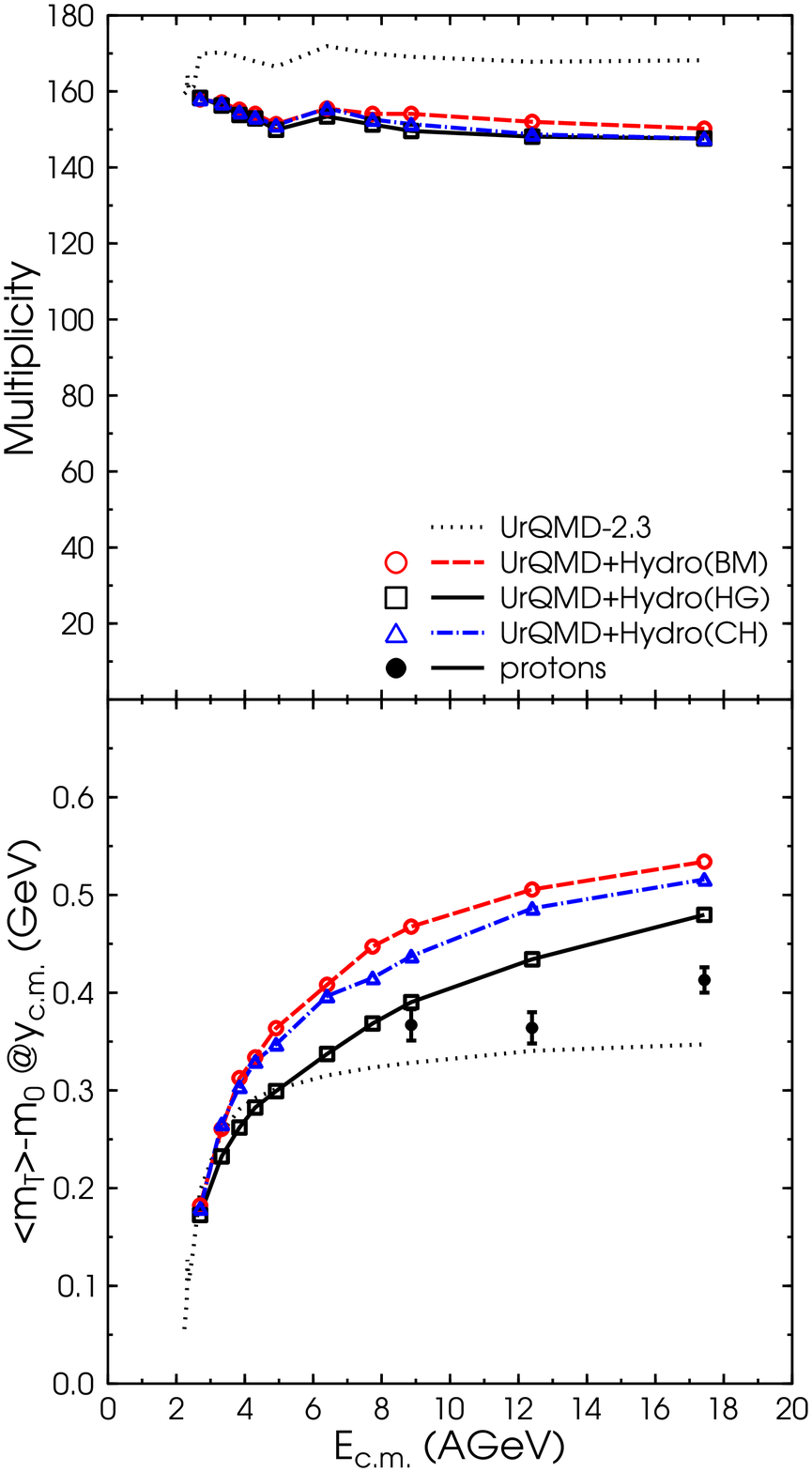}
\caption[EoS dependence of the $\langle m_T \rangle$ excitation function of pions and protons]{\label{fig_eospi} The multiplicity (4$\pi$, top) and the $\langle m_T \rangle$ (bottom) excitation function for pions (left) and protons (right) in central ($b<3.4$ fm) Au+Au/Pb+Pb collisions at $E_{\rm lab}=2-160 ~A$GeV is shown. The lines depict hybrid model calculations with different equations of state and the pure transport calculation in comparison to the experimental data (symbols)\cite{:2007fe,Afanasiev:2002mx,Ahle:1999uy,Anticic:2004yj}.}
\end{figure}

Fig. \ref{fig_eospi} shows the multiplicity and the mean transverse mass excitation functions for pions and protons. The yields are reduced in the hybrid model compared to the pure transport calculation because of entropy conservation during the ideal hydrodynamic evolution. The changes in the EoS do not affect the multiplicities. The mean transverse mass that is more sensitive to the pressure in the transverse plane is changed. For pions the chiral EoS gives similar results as the hadron gas calculation while the bag model EoS decreases the mean transverse mass at high energies as it is expected for a first order phase transition. The pure transport calculation reproduces the flattening in the intermediate energy regime best.

For the protons Fig. \ref{fig_eospi} (right) the opposite behaviour can be observed. In this case, the BM EoS leads to a higher transverse mass than the chiral EoS which is still higher than the hadronic calculation. UrQMD-2.3 shows the strongest flattening again. The high baryon density regions, where most of the protons are produced at hydrodynamic freeze-out, are perhaps more sensitive to the early stage of the collision where the BM EoS exhibits a higher pressure (in the QGP phase) and the softening due to the mixed phase is only reflected in the mean transverse mass of mesons. The net baryon density is explicitly propagated in the hydrodynamic evolution. Therefore, the final distribution of the baryo-chemical potential at the transition from hydrodynamics to the hadronic cascade reflects the dynamics during the evolution and is sensitive to the initial stopping. The mesons are only influenced by the temperature distribution that mainly depends on the energy density distribution.  

\begin{figure}[t]
\includegraphics[width=0.5\textwidth]{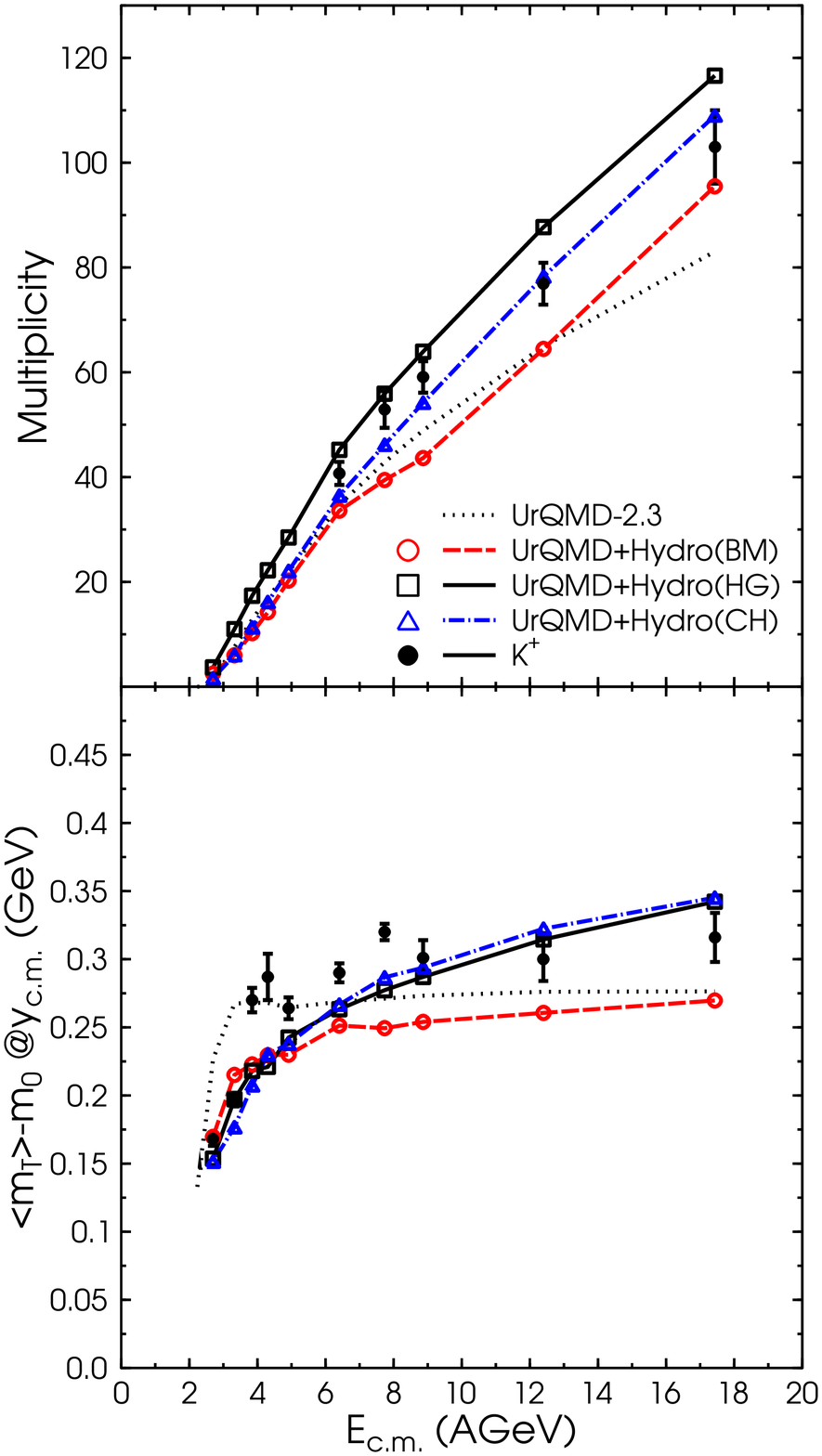}
\includegraphics[width=0.5\textwidth]{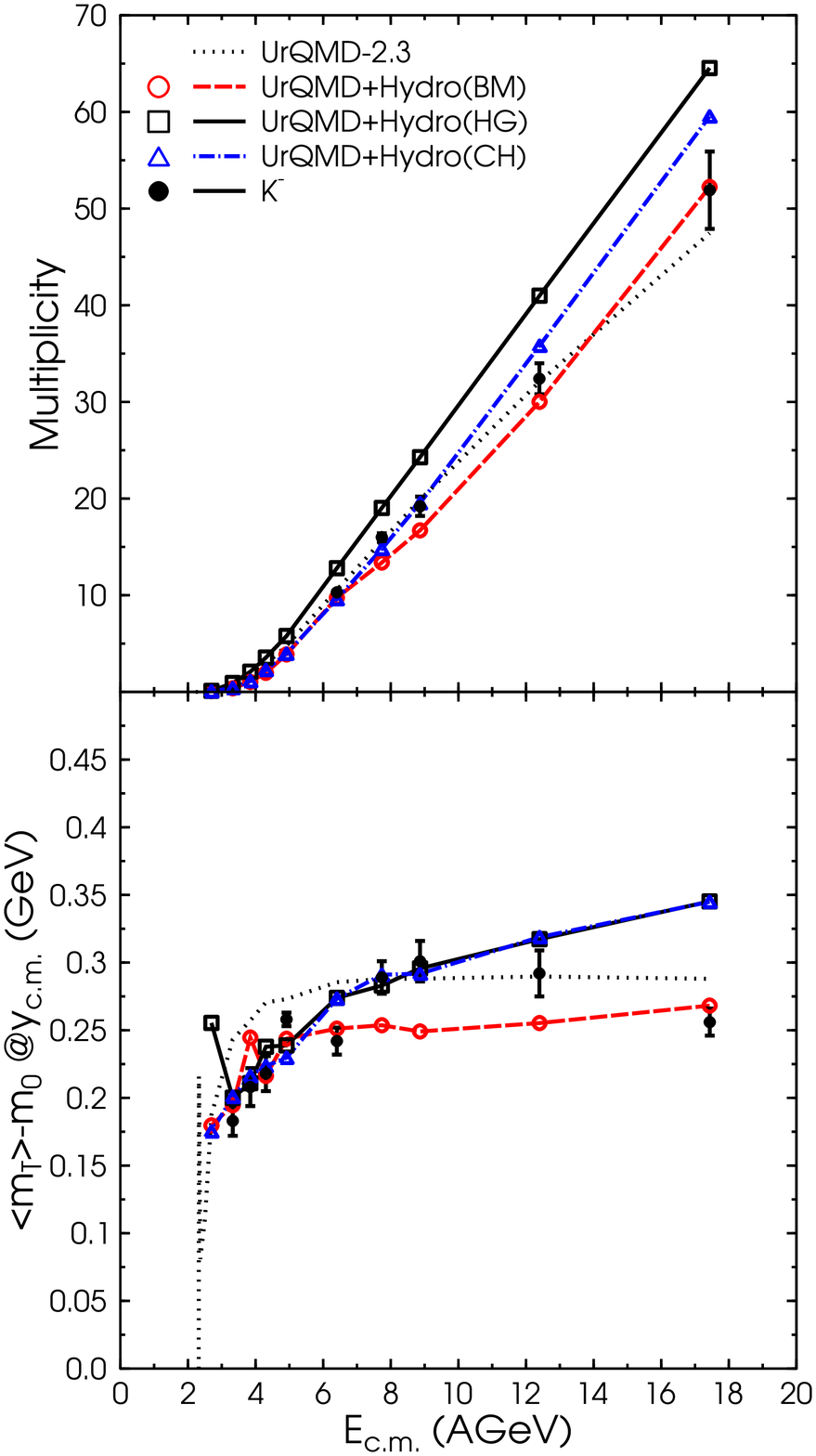}
\caption[EoS dependence of the $\langle m_T \rangle$ excitation function of kaons]{\label{fig_eoskp}The multiplicity (4$\pi$, top) and the $\langle m_T \rangle$ (bottom) excitation function for positively/negatively (left/right) charged kaons in central ($b<3.4$ fm) Au+Au/Pb+Pb collisions at $E_{\rm lab}=2-160 ~A$GeV is shown. The lines depict hybrid model calculations with different equations of state and the pure transport calculation in comparison to the experimental data (symbols)\cite{:2007fe,Afanasiev:2002mx,Ahle:1999uy}.}
\end{figure}

Fig. \ref{fig_eoskp} shows the results for different EoS for positively and negatively charged kaons. Since the qualitative results are the same in both cases we will refer to kaons in the following without distinguishing the charges. The kaon multiplicities are higher in the hybrid model calculation as compared to the pure transport simulation, because strange particles are produced according to thermal distributions during the Cooper-Frye transition. The string and resonance dynamics in UrQMD-2.3 lead to an underestimation of strange particle yields while the mean transverse mass excitation function flattens at high AGS energies due to non-equilibrium effects. The multiplicities and the mean transverse mass are highest in the hadron gas hybrid calculation because the hadronic EoS is the stiffest EoS and can not be ``softened'' by non-equilibrium effects. Here also, the transverse expansion is more violent since the pressure and its gradients are large during the whole hydrodynamic evolution. The chiral EoS leads to a small decrease in the yields, but leaves the $\langle m_T \rangle$ excitation function essentially unchanged. Employing the chiral EoS the temperatures at the transition from hydrodynamics to transport are a bit lower than in the purely hadronic case, but the kaons acquire approximately the same transverse momentum during the evolution. The chiral phase transition exhibits only a small latent heat and therfore the pressure gradients are not affected that much. The BM calculation with a strong first order phase transition produces lower kaon multiplicites than the other hybrid model calculations, at low SPS energies even less than the pure transport calculation. The transition temperatures in this case are the lowest due to the long duration of the hydrodynamic stage. For the kaons the flattening of the mean transverse mass excitation function due to the softening of the EoS is best visible. The mean transverse mass values are even lower than in the nonequilibrium transport calculation. 

The multiplicities are reasonable well reproduced in all the different scenarios, but the mean transverse mass excitation function reflects the different transverse pressure gradients due to the underlying EoS. The experimentally measured step-like behaviour can either be attributed to a first order phase transition with a large latent heat or a softening of the EoS due to hadronic non-equilibrium effects.

To bring the calculations to a better agreement four options with
different (dis)advantages may be explored:
\begin{enumerate}
\item
Inclusion of potential interactions in the initial and/or final stage
of the transport calculations. This might help to lower the mean
transverse momentum of hadrons that are attracted by the potential
interaction. However, it would lead to a higher pressure (and therefore
$\langle m_T \rangle$) especially for the protons which are already overestimated by the
existing approach.

\item
Changing the equation of state. Another possibility is to increase
the latent heat in the bag model EoS to get a stronger effect of the
softening. This would eventually reduce the $\langle m_T \rangle$ of all hadrons bringing
it nearer to the data. However, it would lead to a further increase of
the expansion times due the long timespan that the system spends
in the mixed phase. This lifetime increase seems not to be supported by
observables, like HBT radii, that are sensitive to the lifetime of the
fireball.

\item
Modifications of the freeze-out. The freeze-out prescription is a
further crucial ingredient as it has been shown in the last Section. In
the present approach, a transition criterion on a constant hypersurface
and the Cooper-Frye-approach is used. However, one might also think of a
continuous emission approach to the decoupling problem without involving
the Cooper-Frye prescription. Here particle emission is integrated over
various stages of the reaction and might result in a change of the
transverse momentum spectra compared to the CF approach. These models
have been explored e.g. in \cite{Grassi:1994nf,Knoll:2008sc}.

\item
Viscosities. All calculations with the ideal hydrodynamics
intermediate stage overpredict the $\langle m_T \rangle$, however, the non-equilibrium
approach UrQMD is generally on the lower side of the data, non-equilibrum effects seem to play an important role. Thus, adding
additional viscosity during the hydrodynamic evolution might be the most
promising idea to improve the overall agreement with the experimental
data.

\end{enumerate}
\section[Summary]{Summary}
\label{sum}

The multiplicity and mean transverse mass excitation function has been calculated for pions, protons and kaons in the energy range from $E_{\rm lab}=2-160A~$GeV. An integrated Boltzmann+hydrodynamics approach has been used which allows for a systematic investigation of different effects. The initial conditions and the freeze-out are uniquely fixed and generated via the UrQMD transport approach and different equations of state are explored. 

First, we have investigated the dependence of the results on a change of the freeze-out prescription. It was observed that the different freeze-out procedures have almost as much influence on the mean transverse mass excitation function as the EoS. A comparison to the available data suggests that a gradual transition from hydrodynamics to the transport simulation at an energy density of 5$\epsilon_0$ provides the best description of the data. The experimentally observed step-like behaviour of the mean transverse mass excitation function is only reproduced, if a first order phase transition with a large latent heat is applied or the EoS is effectively softened due to non-equilibrium effects in the hadronic transport calculation. 

To distinguish between these two scenarios it is useful to investigate different observables that are more sensitive to the mean free path like e.g. the elliptic flow or the expansion times which manifest themselves in the $R_O/R_S$ ratio. The flow is expected to be larger in a hydrodynamic calculation than in a transport calculation due to the higher pressure gradients in the early stage of the collision. 

\section*{Acknowledgements}
This work was supported by BMBF and GSI. The computational resources were provided by the Frankfurt 
Center for Scientific Computing (CSC). We would like to thank Stefan Schramm for helpful discussions. The authors thank Dirk Rischke for providing the 1f-hydrodynamics code. H. Petersen thanks the Deutsche Telekom Stiftung for the scholarship and the Helmholtz Research School on Quark Matter Studies for additional support. This work was supported by the Helmholtz International Center for FAIR within the framework of the LOEWE program launched by the State of Hesse.

\end{document}